\documentclass{moriond}

\def\Journal#1#2#3#4{{#1} {\bf #2}, #3 (#4)}

\def\PLB{{\em Phys. Lett.}  B}

\def\PRD{{\em Phys. Rev.} D}

\def\JINST{\em JINST}
\def\JHEP{\em JHEP}

\def\ttbar{\ensuremath{\mathrm{t}\bar{\mathrm{t}}}}
\def\dphi{\ensuremath{\left|\Delta \phi_{\ell\ell}\right|}}
\def\deta{\ensuremath{\left|\Delta\eta\right|}}
\def\fSM{\ensuremath{f_{\mathrm{SM}}}}
\def\Powhegvtwo{{\textsc{Powhegv2}}}
\def\ctgl{\ensuremath{C_\mathrm{tG}/\Lambda^{2}}}

\def\be{\begin{equation}}
\def\ee{\end{equation}}
\def\bea{\begin{eqnarray}}
\def\eea{\end{eqnarray}}

\providecommand{\eqn}{Eq.}
\providecommand{\fig}{Fig.}

\providecommand{\secn}{Section}

\newcommand{\Photo}{}

\begin{document}
\vspace*{4cm}
\title{Spin correlations in top physics at ATLAS and CMS in Run 2}

\author{Jacob Linacre, {on behalf of the ATLAS and CMS Collaborations}}

%\address{STFC RAL, {jacob.linacre@gmail.com}}
\address{STFC Rutherford Appleton Laboratory, Harwell Campus, Didcot OX11 0QX, England}

\maketitle\abstracts{
Measurements of \ttbar\ spin correlations are presented in events with top quarks produced in $\mathrm{pp}$ collisions at the LHC. 
The data correspond to an integrated luminosity of $36\:\mathrm{fb^{-1}}$ at $\sqrt{s}=13\:\mathrm{TeV}$ collected at both the ATLAS and CMS detectors. 
The spin correlations are measured using the angular distributions of the leptons in dilepton channel \ttbar\ events. 
The spin correlations are probed both directly, using distributions measured in the top quark rest frames that depend only on the top quark spin, and indirectly, using distributions measured in the laboratory frame. 
The distributions are unfolded to the parton level and extrapolated to the full phase space. Some of the laboratory frame distributions are additionally unfolded to the particle level in the fiducial phase space of the ATLAS detector. 
The spin correlation measurements are used to search for new physics in the form of a light top squark or an anomalous top quark chromo-magnetic dipole moment, and stringent constraints are placed in both cases.
}

\section{Introduction}

The large mass of the top quark and its corresponding strong coupling to the Higgs boson suggest a connection between the top quark and the mechanism of electroweak symmetry breaking. 
New physics in this mechanism is likely to modify the spin properties of top quark pair (\ttbar) events from the standard model (SM) expectations, either via underlying direct production modes or from interference effects from new physics at higher mass scales~\cite{zhang}. Furthermore, the top quark is the only quark that decays before hadronising, so the information about its spin is transferred to its decay products undiluted by non-perturbative effects. The charged lepton in top quark decay is a perfect spin analyser~\cite{Brandenburg2002235}, meaning its angular distribution retains the maximum amount of information about the top quark spin. Top quark spin measurements in dilepton \ttbar\ events therefore provide an ideal laboratory to test perturbative QCD and probe for new physics.

At the LHC \ttbar\ production proceeds primarily via the strong interaction (mostly $\mathrm{gg} \to \ttbar$), which at the leading order (LO) produces unpolarised top quarks. A small top quark polarisation, measured relative to the direction of the recoiling top quark, arises when including electroweak corrections,
while a small polarisation transverse to the scattering plane arises from absorptive terms at one loop (both ${<}\,1\%$~\cite{Bernreuther2015}).
However, the spins of the top quarks and antiquarks are strongly
correlated.
The SM predicts a rich structure of spin correlations, where
the configuration of quark-antiquark spins is dependent on
both the initial state and the top quark production kinematics~\cite{newtwist}.
The analysed LHC $\mathrm{pp}$ collision data corresponds to an integrated luminosity of $36\:\mathrm{fb^{-1}}$ at $\sqrt{s}=13\:\mathrm{TeV}$ collected at both the ATLAS~\cite{ATLASdet} and CMS~\cite{CMSdet} detectors.

\section{Indirect measurements of spin correlations \label{sec:indirect}}

\begin{figure}
\begin{minipage}{0.325\linewidth}
\centerline{\includegraphics[width=1.0\linewidth]{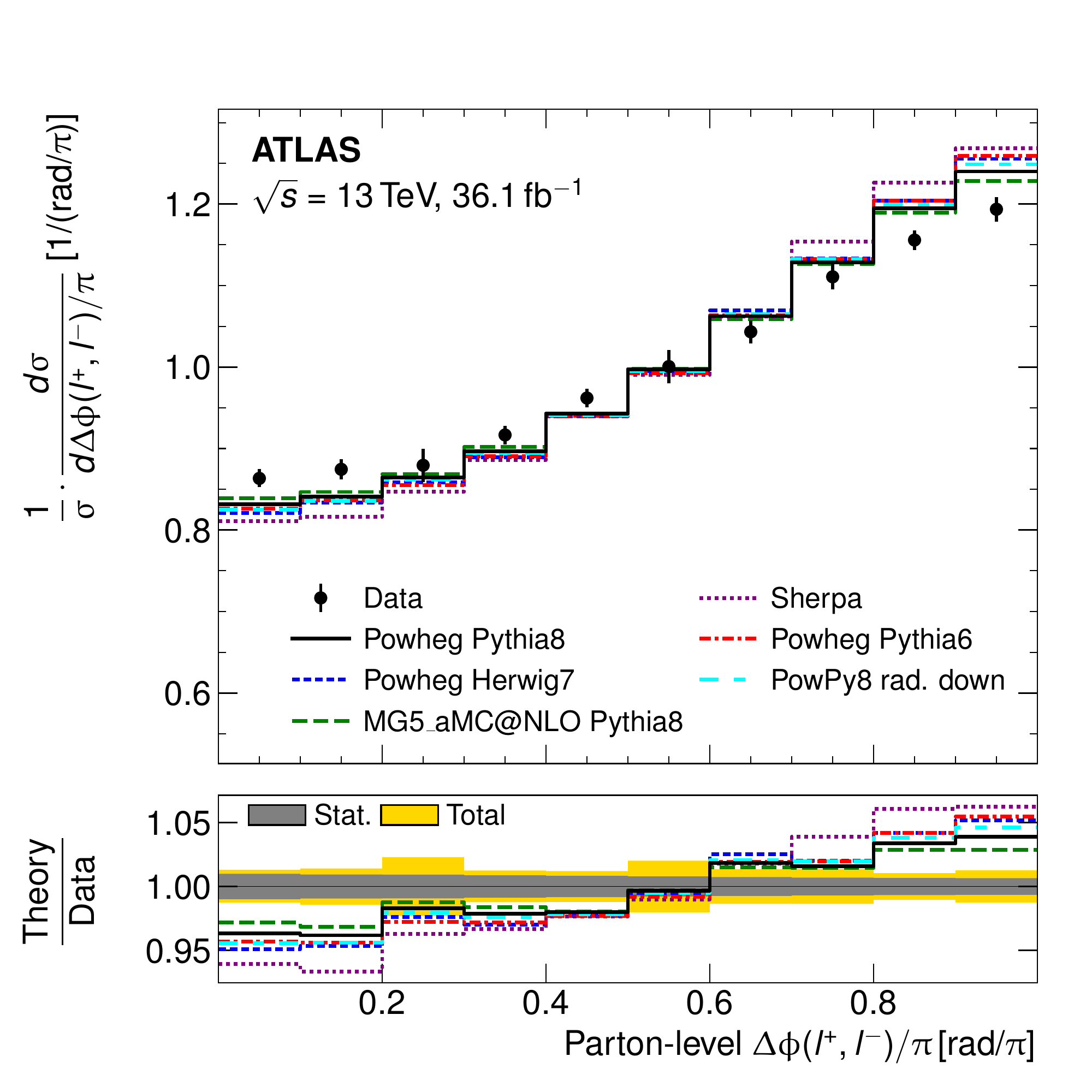}}
\end{minipage}
\hfill
\begin{minipage}{0.325\linewidth}
\centerline{\includegraphics[width=1.0\linewidth]{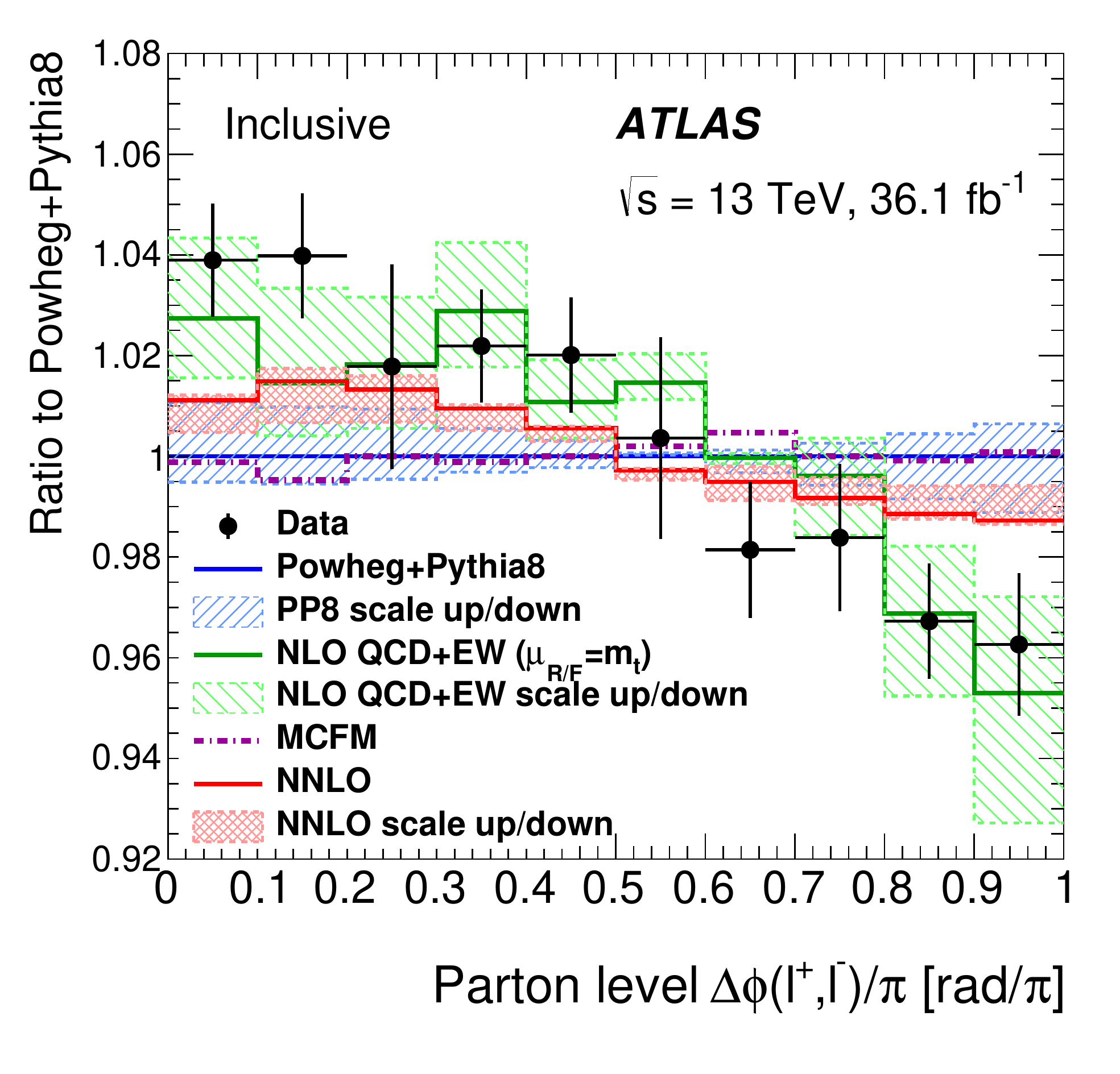}}
\end{minipage}
\hfill
\begin{minipage}{0.335\linewidth}
%\centerline{\includegraphics[width=1.0\linewidth]{fig_11.pdf}}
\centerline{\includegraphics[width=1.0\linewidth]{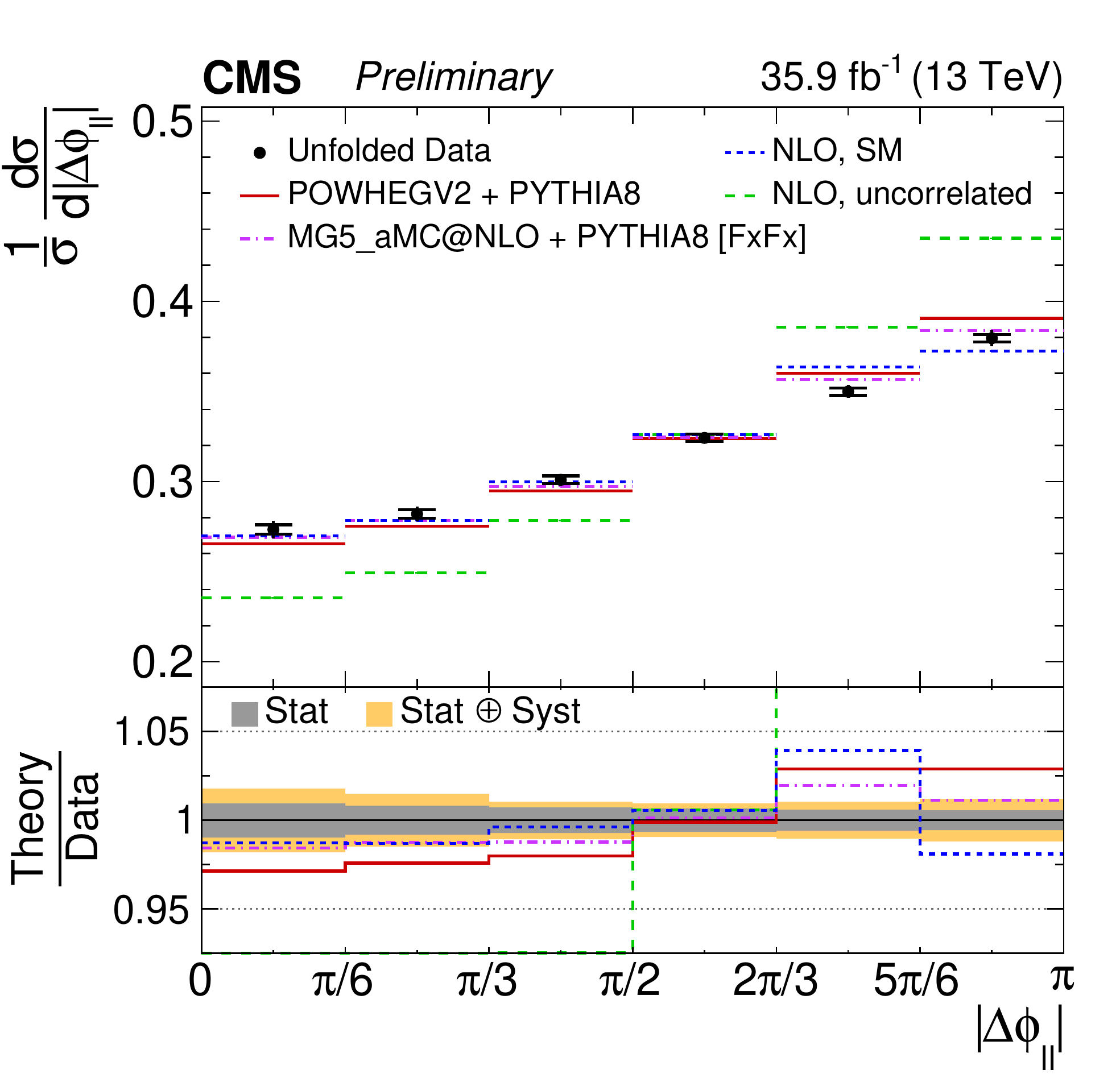}}
\end{minipage}
\caption[]{Parton-level \dphi\ distributions measured by ATLAS~\cite{ATLASresult} (left) and CMS~\cite{CMSresult} (right), compared with various predictions. In the centre, the ATLAS measurement is compared with fixed-order calculations~\cite{Bernreuther2015,NNLO}.}
\label{fig:indirect}
\end{figure}

The correlation between the top quark and antiquark spins induces a tendency towards alignment of the decay angles of the daughter leptons. This correlation is most strongly observed in the transverse plane, and is retained in the separation in azimuthal angle of the two leptons in the laboratory frame (\dphi). The approximately back-to-back configuration of the parent top quark and antiquark results in preference for large \dphi, while the spin correlations generate a relative enhancement of ${\approx}\,15\%$ at low \dphi\ (see \fig~\ref{fig:indirect}, right). The \dphi\ distribution can be very precisely reconstructed, owing to the excellent experimental angular resolution of the lepton measurements, and it is therefore an important probe of deviations from the SM.

\begin{figure}
\begin{minipage}{0.341\linewidth}
\centerline{\includegraphics[width=1.0\linewidth]{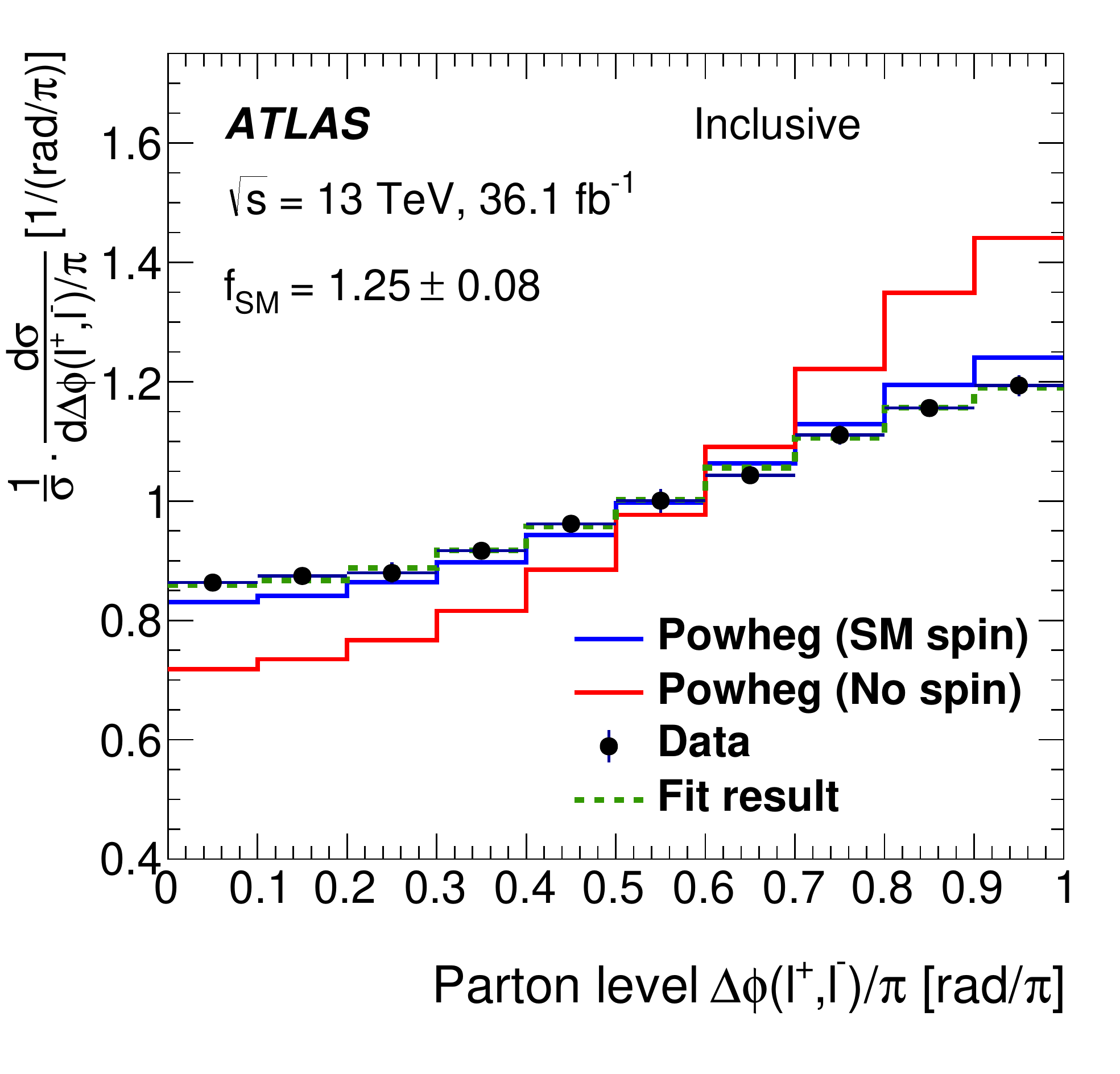}}
\end{minipage}
\hfill
\begin{minipage}{0.314\linewidth}
\centerline{\includegraphics[width=1.0\linewidth]{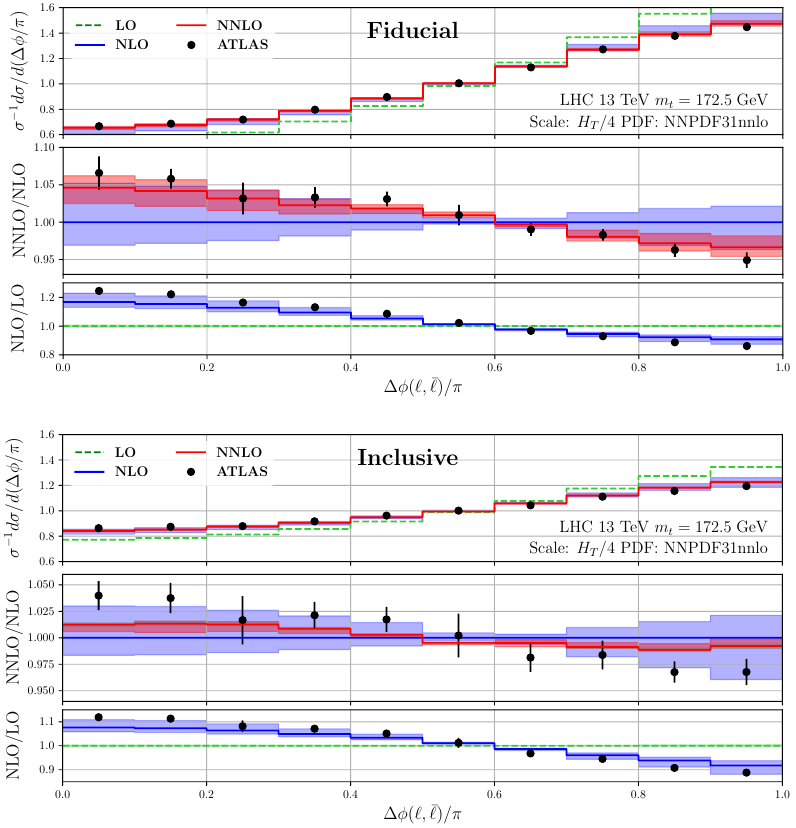}}
\end{minipage}
\hfill
\begin{minipage}{0.327\linewidth}
\centerline{\includegraphics[width=1.0\linewidth]{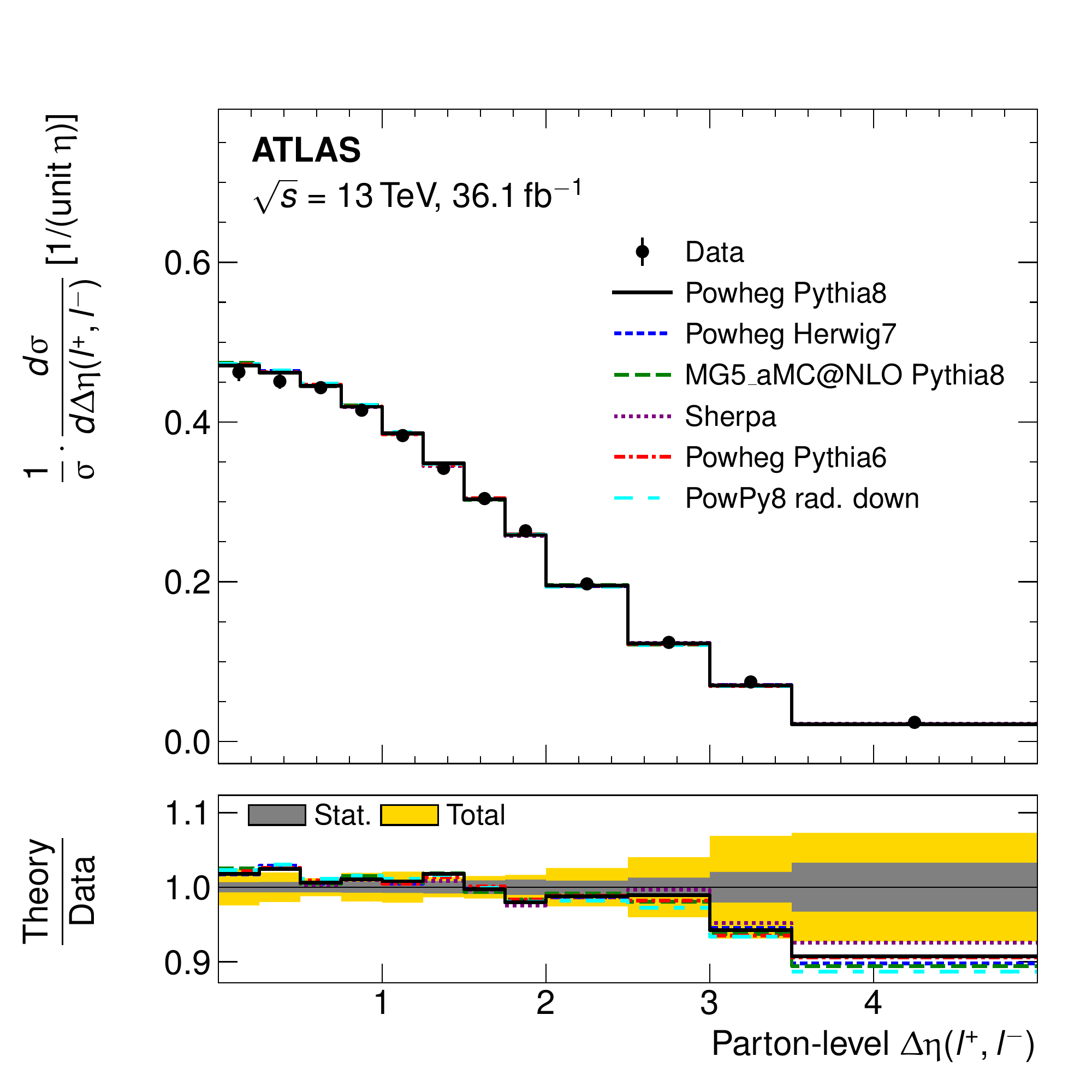}}
\end{minipage}
\caption[]{Left: fit of ATLAS parton-level \dphi\ distribution~\cite{ATLASresult} to a combination of correlated and uncorrelated templates, to measure \fSM. Centre: NNLO QCD calculations of the \dphi\ distribution in the fiducial and inclusive phase space~\cite{NNLO}. Right: parton-level \deta\ distribution measured by ATLAS~\cite{ATLASresult}. This observable has minimal sensitivity to spin correlations, but is sensitive to new physics in \ttbar\ production (see \secn~\ref{sec:susy}).}
\label{fig:indirect2}
\end{figure}

The results from ATLAS~\cite{ATLASresult} and CMS~\cite{CMSresult}, unfolded to the parton level and extrapolated to the full phase space, are shown in \fig~\ref{fig:indirect}, along with various predictions from simulation~\cite{powheg,mg5aMC} and fixed-order calculations~\cite{Bernreuther2015,NNLO}. There is a clear preference of the data for the predictions including spin correlations, and the ATLAS experiment uses the predictions of the next-to-LO (NLO) \Powhegvtwo\ generator~\cite{powheg} to fit the strength of the spin correlations as a fraction of the SM prediction, \fSM\ (\fig~\ref{fig:indirect2}, left). The result is $\fSM=1.25 \pm 0.02\,\mathrm{(stat)} \pm 0.06\,\mathrm{(syst)} \pm 0.04\,\mathrm{(theo)}$, where the combined uncertainty is $\pm 0.08$, suggesting that the observed spin correlations are $3.2\sigma$ stronger than those predicted by the SM.
The dominant source of systematic uncertainty is the choice of renormalisation and factorisation scales ($\mu_\mathrm{R}$ and $\mu_\mathrm{F}$) and the amount of initial- and final-state radiation (ISR and FSR) in the generated \ttbar\ events.
The discrepancy remains when ATLAS repeats the measurement, limiting the extrapolation of the measured distribution to the fiducial phase space of the detector, and a similar discrepancy is observed in the \dphi\ disribution measured by CMS in the full phase space (\fig~\ref{fig:indirect}, right). A possible resolution appeared recently in the first full next-to-NLO (NNLO) QCD calculations of the \dphi\ distribution~\cite{NNLO}. The corrections are found to be small in the full phase space, but in a fiducial phase space similar to that of the ATLAS and CMS detectors they are large enough to account for much of the observed discrepancy (\fig~\ref{fig:indirect2}, centre). Since the ratio of the full and fiducial phase space differential cross sections is used in the extrapolation of the measured distribution to the full phase space, this effect can account for the observed discrepancy in the full phase space as well.
%although NLOW...
%and \fig~\ref{fig:indirect2}, left
%The unfolded \deta\ distribution measured by ATLAS is shown in \fig~\ref{fig:indirect2}. It has minimal sensitivity to spin correlations, but \secn

\section{Direct measurements of spin correlations \label{sec:direct}}

\begin{figure}
\begin{minipage}{0.305\linewidth}
\centerline{\includegraphics[width=1.0\linewidth]{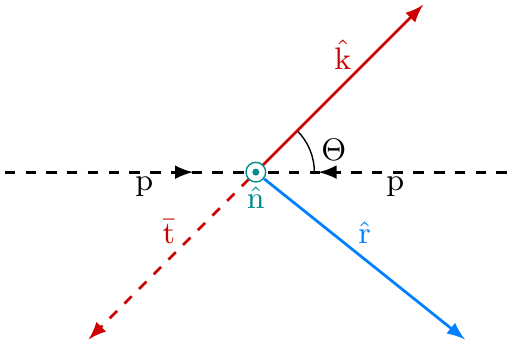}}
\end{minipage}
\hfill
\begin{minipage}{0.34\linewidth}
\centerline{\includegraphics[width=1.0\linewidth]{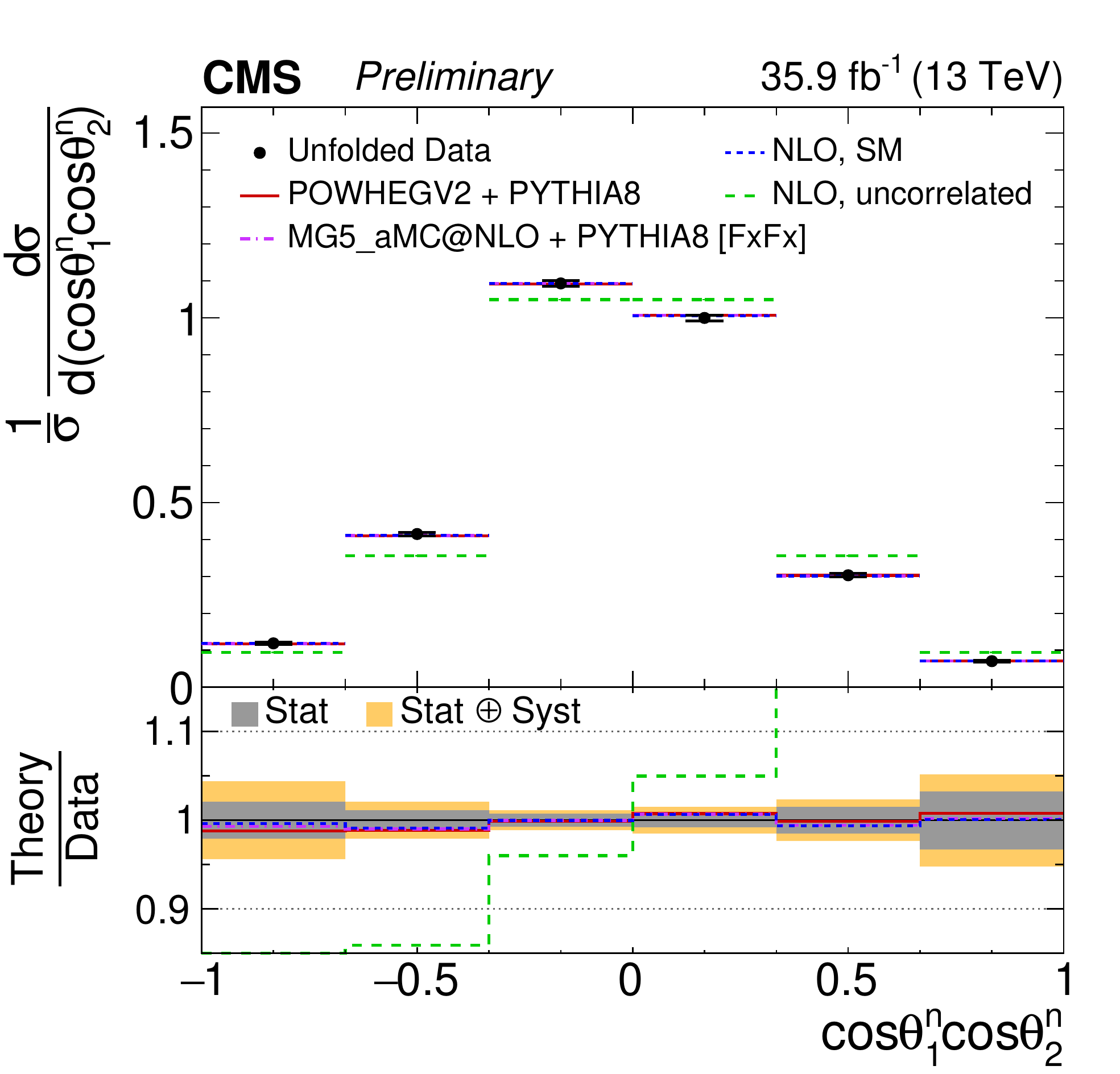}}
%\centerline{\includegraphics[width=1.0\linewidth]{UnfoldedResultsNorm_ll_cHel.pdf}}
\end{minipage}
\hfill
\begin{minipage}{0.34\linewidth}
\centerline{\includegraphics[width=1.0\linewidth]{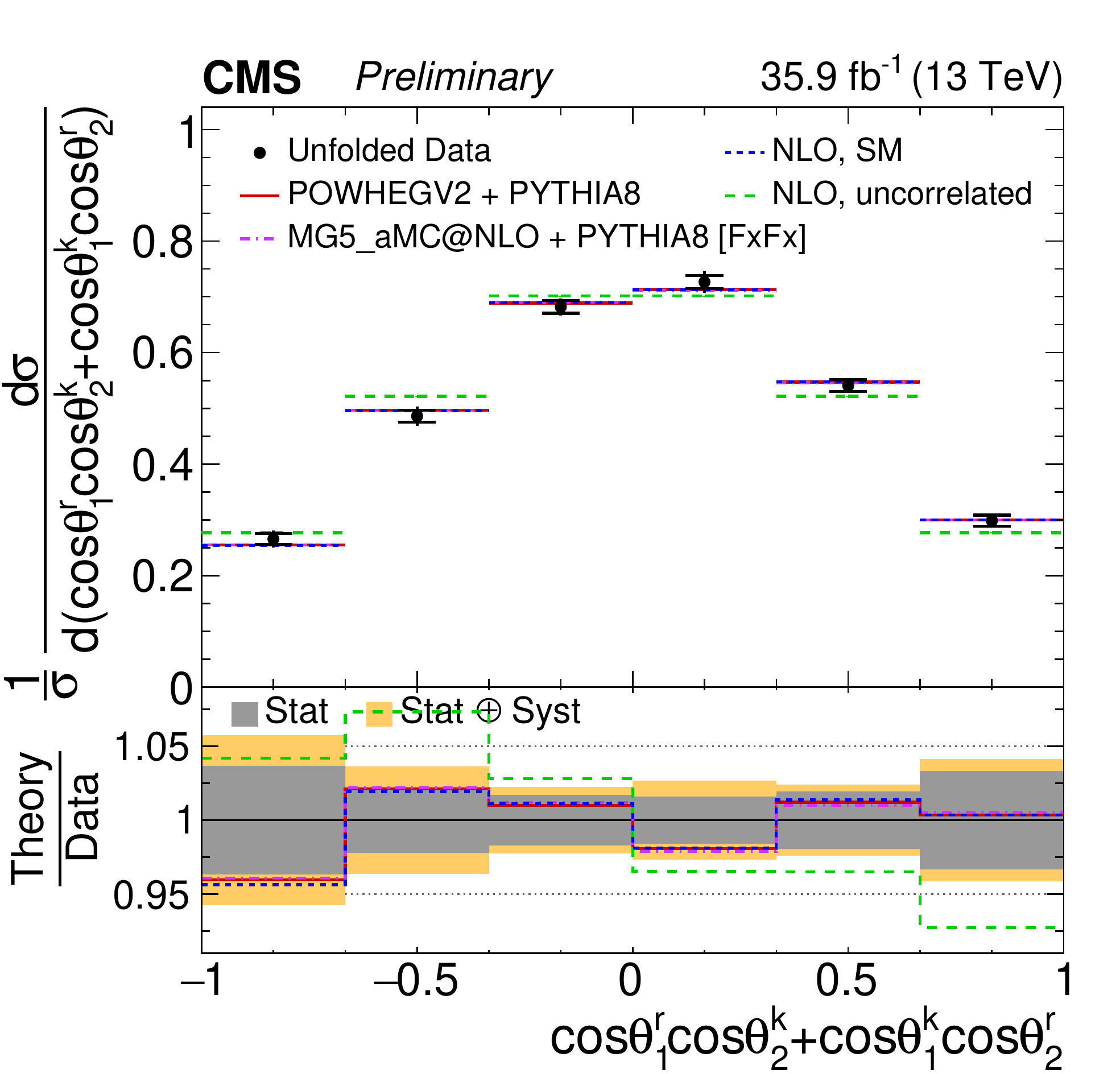}}
\end{minipage}
\caption[]{Left: coordinate system used for the top quark spin measurements, shown in the $\mathrm{pp} \to \ttbar$ scattering plane. Centre: distribution representing the correlation between the lepton directions both measured with respect to the $\hat{n}$ axis, probing $C_{nn}$~\cite{CMSresult}. Right: distribution representing the correlation between the lepton directions, one measured with respect to the $\hat{r}$ axis and the other measured with respect to the $\hat{k}$ axis, probing $C_{rk}+C_{kr}$~\cite{CMSresult}.}
\label{fig:direct}
\end{figure}

Using the lepton directions measured in their parent top quark rest frames as proxies for the top quark spins, all of the spin-dependent parts of the \ttbar\ production density matrix can be probed~\cite{Bernreuther2015}:
\begin{equation}
\frac{1}{\sigma}\frac{\mathrm{d}\sigma}{\mathrm{d}\Omega_1\mathrm{d}\Omega_2} = \frac{1}{(4\pi)^2} \left( 1 + \mathbf{B_1} \cdot \hat{\ell}_1 + \mathbf{B_2} \cdot \hat{\ell}_2 - \hat{\ell}_1 \cdot C \cdot \hat{\ell}_2  \right),
\label{eq:fulldist}
\end{equation}
where $\mathbf{{B}_{1,2}}$ are three-dimensional vectors that characterise the degree of top quark or antiquark polarisation in each direction, 
and ${C}$ is a $3\times3$ matrix that characterises the correlation between the top quark and antiquark spins.
The spin is measured using a basis, illustrated in \fig~\ref{fig:direct}, chosen such that the $B_{1,2}^i$ and $C_{ij}$ coefficients (the elements of the $\mathbf{B_1}$ and $\mathbf{B_2}$ vectors and of the $C$ matrix) have definite properties with respect to discrete symmetries such as C and P~\cite{Bernreuther2015}.

%Each element of the matrix ${C}$ can be probed by a one-dimensional angular distribution measured at the parton level, of the form
Each coefficient is probed by the CMS experiment by measuring~\cite{CMSresult} a one-dimensional angular distribution derived from \eqn~\ref{eq:fulldist}, which for each coefficient has the form
\begin{equation}
%\frac{1}{\sigma}\frac{\mathrm{d}\sigma}{\mathrm{d}x} = \frac{1}{2} \left( 1 + C_{ij} \, x \right)  f(x).
\frac{1}{\sigma}\frac{\mathrm{d}\sigma}{\mathrm{d}x} = \frac{1}{2} \left( 1 + \mathrm{Coefficient} \times x \right)  f(x).
\label{eq:simpdist}
\end{equation}
These distributions are sensitive only to the top quark spin (independent of the top quark kinematics), and the measurements are therefore less affected by theoretial uncertainties.
However, compared to the indirect measurements the statistical precision of the measurements is diluted by the poor resolution of the top quark momentum reconstruction, caused largely by the presence of two neutrinos and ambiguities in the assignments of measured jets to the b~quarks from top quark decay.
The measured normalised differential cross sections are unfolded to the parton level and extrapolated to the full phase space. The known functional forms of \eqn~\ref{eq:simpdist}, which are unaffected by new physics in \ttbar\ production, are used to construct an unbiased regularisation of the unfolding.
In addition to full statistical and systematic covariance matrices for each measured distribution, matrices are calculated for the set of all measured bins, allowing constraints to be placed using several measured distributions simultaneously (an example can be seen in \fig~\ref{fig:interp}).
A selection of results for distributions probing the $C$ matrix is shown in \fig~\ref{fig:direct}.  
% only to spin correlations

\begin{figure}
\begin{minipage}{0.495\linewidth}
\centerline{\includegraphics[width=1.0\linewidth]{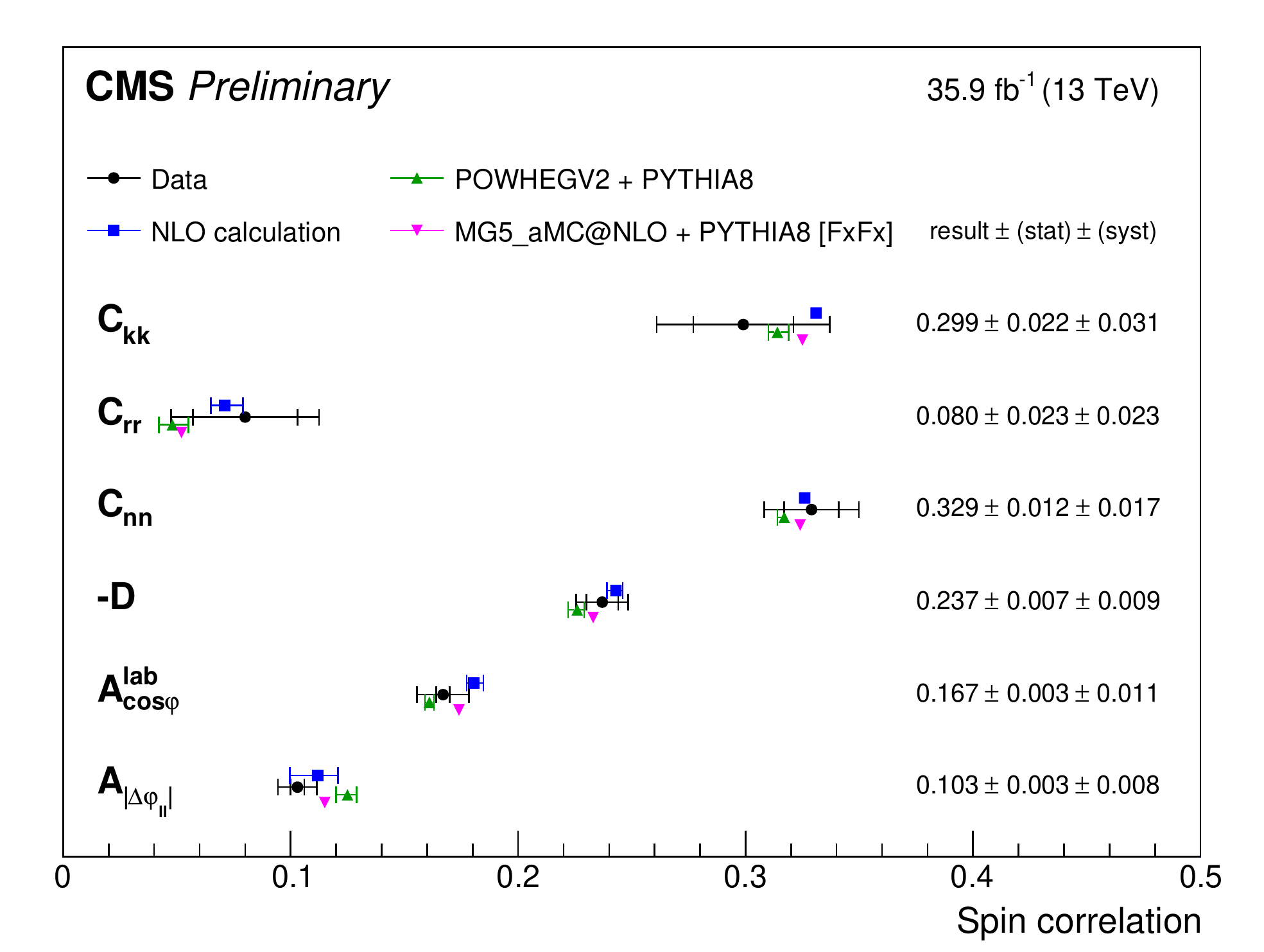}}
\end{minipage}
\hfill
\begin{minipage}{0.495\linewidth}
\centerline{\includegraphics[width=1.0\linewidth]{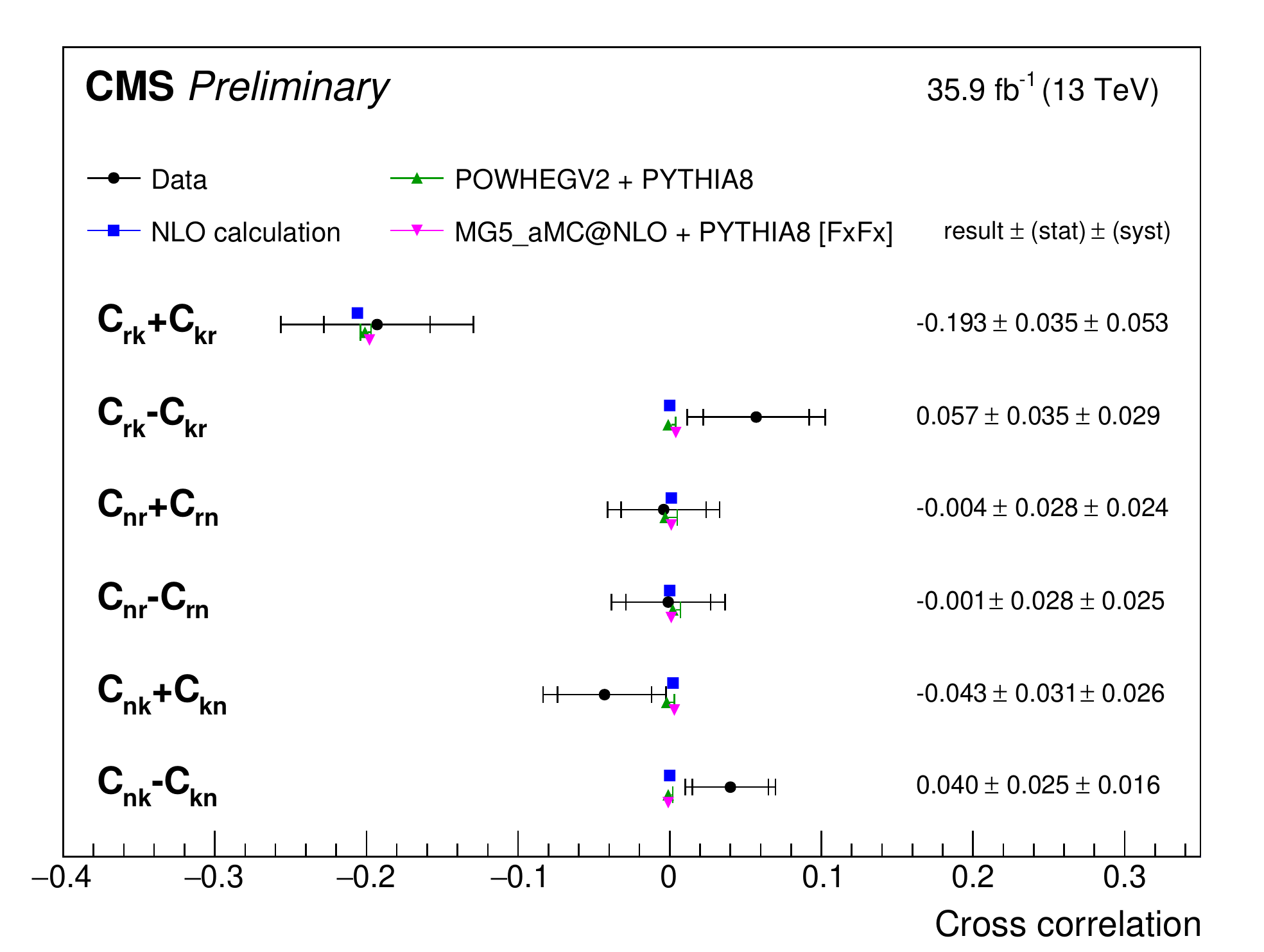}}
%\centerline{\includegraphics[width=1.0\linewidth]{UnfoldedResultsNorm_ll_cHel.pdf}}
\end{minipage}
\caption[]{Summary of measured~\cite{CMSresult} diagonal (left) and off-diagonal (right) coefficients of the matrix $C$.}
\label{fig:direct2}
\end{figure}

From each measured normalised differential cross section CMS extracts the corresponding spin correlation coefficient, and the results are shown in \fig~\ref{fig:direct2}. The top quark polarisation is also measured with respect to each reference axis, but the measurements are all consistent with zero and not sensitive to the small level of top quark polarisation predicted in the SM.
The systematic and statistical uncertainties are of comparable size for most of the measured coefficients, and 
the dominant sources of systematic uncertainty are typically in the jet energy scale and b~quark fragmentation (which affect the reconstruction of the top quark rest frame), the background subtraction, and the \ttbar\ simulation modelling (ISR and FSR, $\mu_\mathrm{R}$ and $\mu_\mathrm{F}$, and top quark $p_{\mathrm T}$).

\begin{figure}
\begin{minipage}{0.44\linewidth}
\centerline{\includegraphics[width=1.0\linewidth]{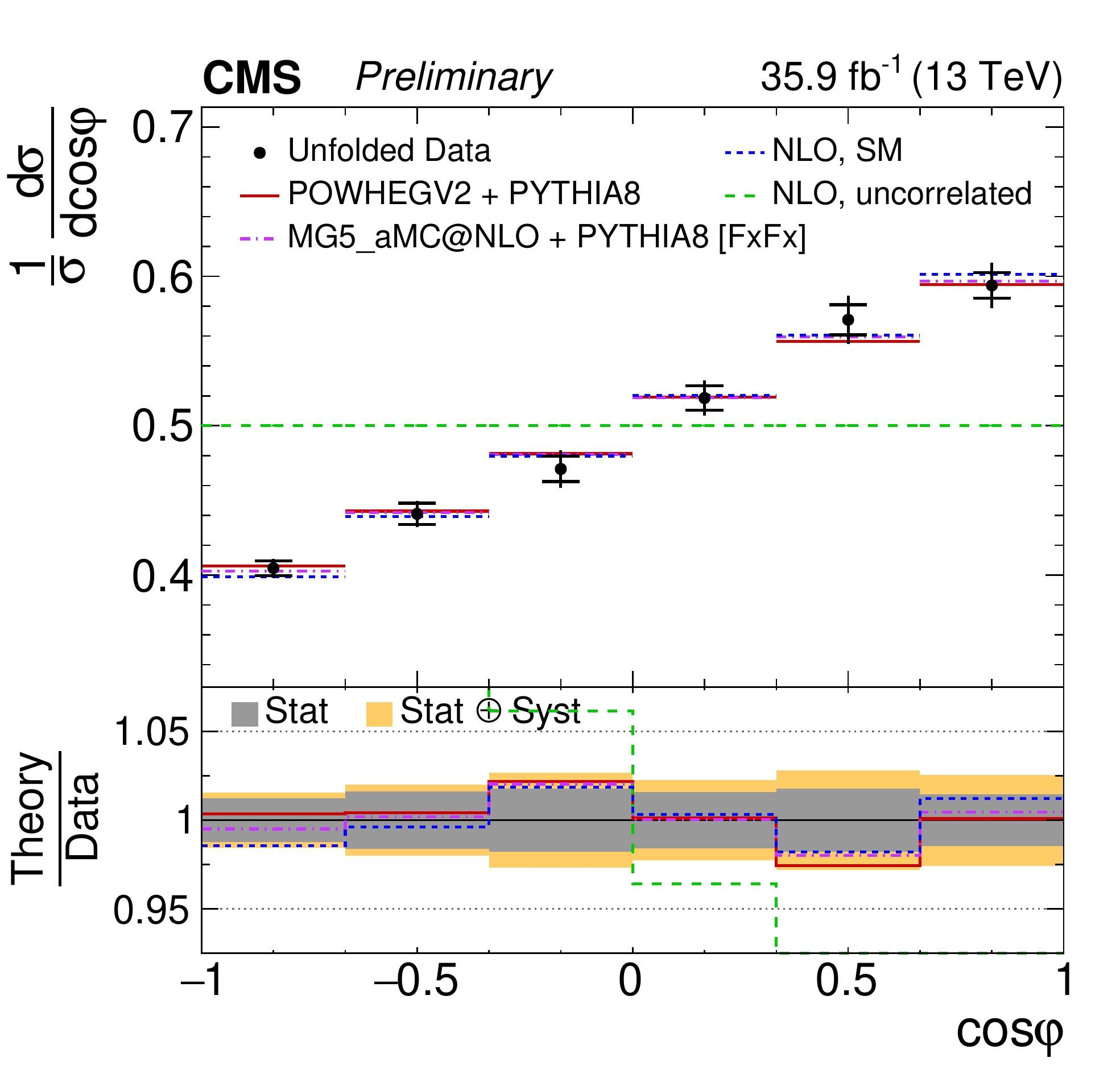}}
\end{minipage}
\hfill
\begin{minipage}{0.55\linewidth}
\centerline{\includegraphics[width=1.0\linewidth]{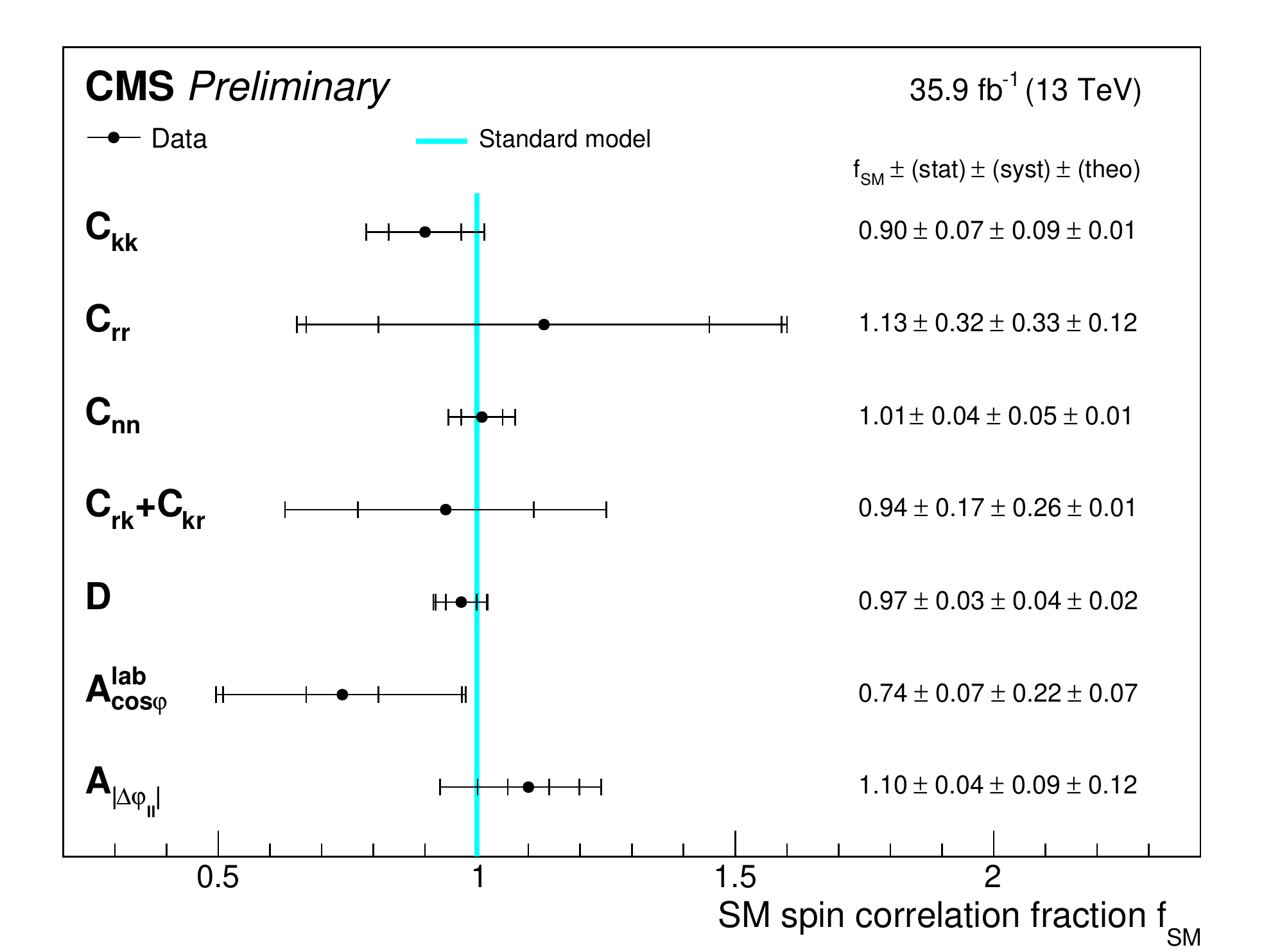}}
\end{minipage}
\caption[]{Left: measured distribution~\cite{CMSresult} of the cosine of the angle between the lepton directions measured in their parent top quark rest frames, sensitive to the spin correlation coefficient $D$ (related to the trace of the $C$ matrix). The prediction without spin correlations (uncorrelated) is strongly disfavoured. Right: summary of \fSM\ results~\cite{CMSresult}.}
\label{fig:direct3}
\end{figure}

The measured coefficients are converted into values of \fSM\ using theoretical predictions 
at NLO in QCD with EW corrections~\cite{Bernreuther2015} (all coefficients are zero in the absence of spin correlations), and the results are shown in \fig~\ref{fig:direct3} (right).
The most precise measurement of spin correlations comes from the distribution of the opening angle between the lepton directions $\hat{\ell}_1$ and $\hat{\ell}_2$ measured in their parent top quark rest frames, $\frac{1}{\sigma}\frac{d\sigma}{d\cos\varphi} = \frac{1}{2}(1-D\cos\varphi)$, where $\cos\varphi=\hat{\ell}_1 \cdot \hat{\ell}_2$ and $D=-\mathrm{tr}(C)/3$. The measured $\cos\varphi$ distribution is shown in \fig~\ref{fig:direct3} (left).
From the $D$ coefficient, CMS measures $\fSM=0.97\pm0.05$. This is the most precise measurement of \fSM\ to date, indicating that the effect of the worse experimental resolution of $\cos\varphi$ compared to \dphi\ is outweighed by the reduced theoretical uncertainties and the stronger dependence on spin correlations of the direct observable.

\section{Interpretation}

\subsection{Anomalous chromomagnetic dipole moment}

The chromomagnetic dipole moment (CMDM) of a
colour-charged particle in colour fields can be defined
by analogy to the magnetic dipole moment of an electrically charged
particle.
In the SM, the
intrinsic spin of the top quark and its colour charge give it a small
CMDM~\cite{Bernreuther2015}. Several beyond-the-SM models
such as two-Higgs-doublet models (e.g., supersymmetry),
technicolor, and top quark compositeness
models~\cite{zhang}, predict an anomalous CMDM,
leading to modifications of the \ttbar\ production rates 
and spin structure.
%and hence to the kinematic
%properties of \ttbar\ events.
%and hence to the leptonic
%angular distributions in \ttbar\ events.

\begin{figure}
\begin{minipage}{0.5088\linewidth}
\centerline{\includegraphics[width=1.0\linewidth]{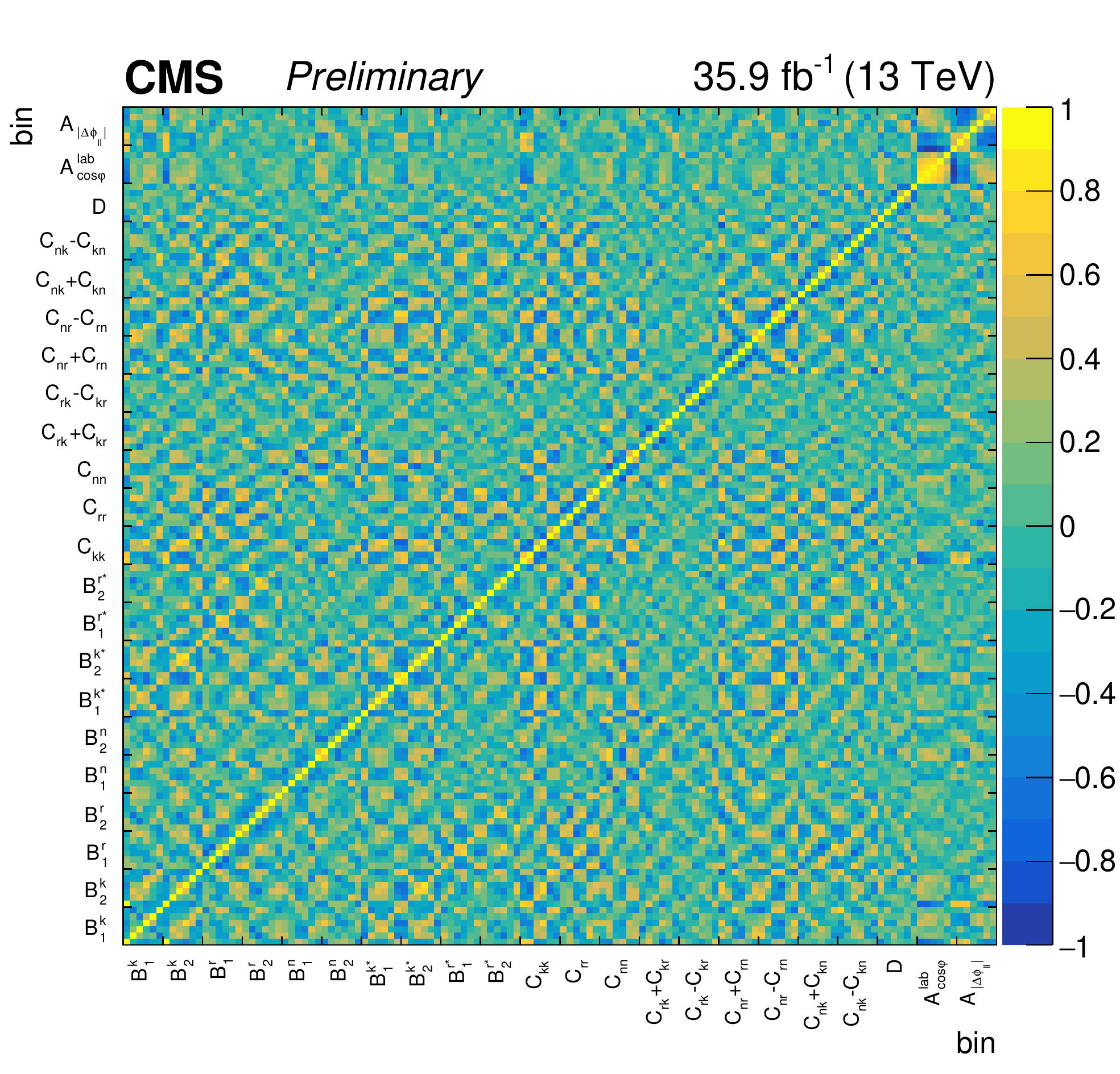}}
\end{minipage}
\hfill
\begin{minipage}{0.475\linewidth}
\centerline{\includegraphics[width=1.0\linewidth]{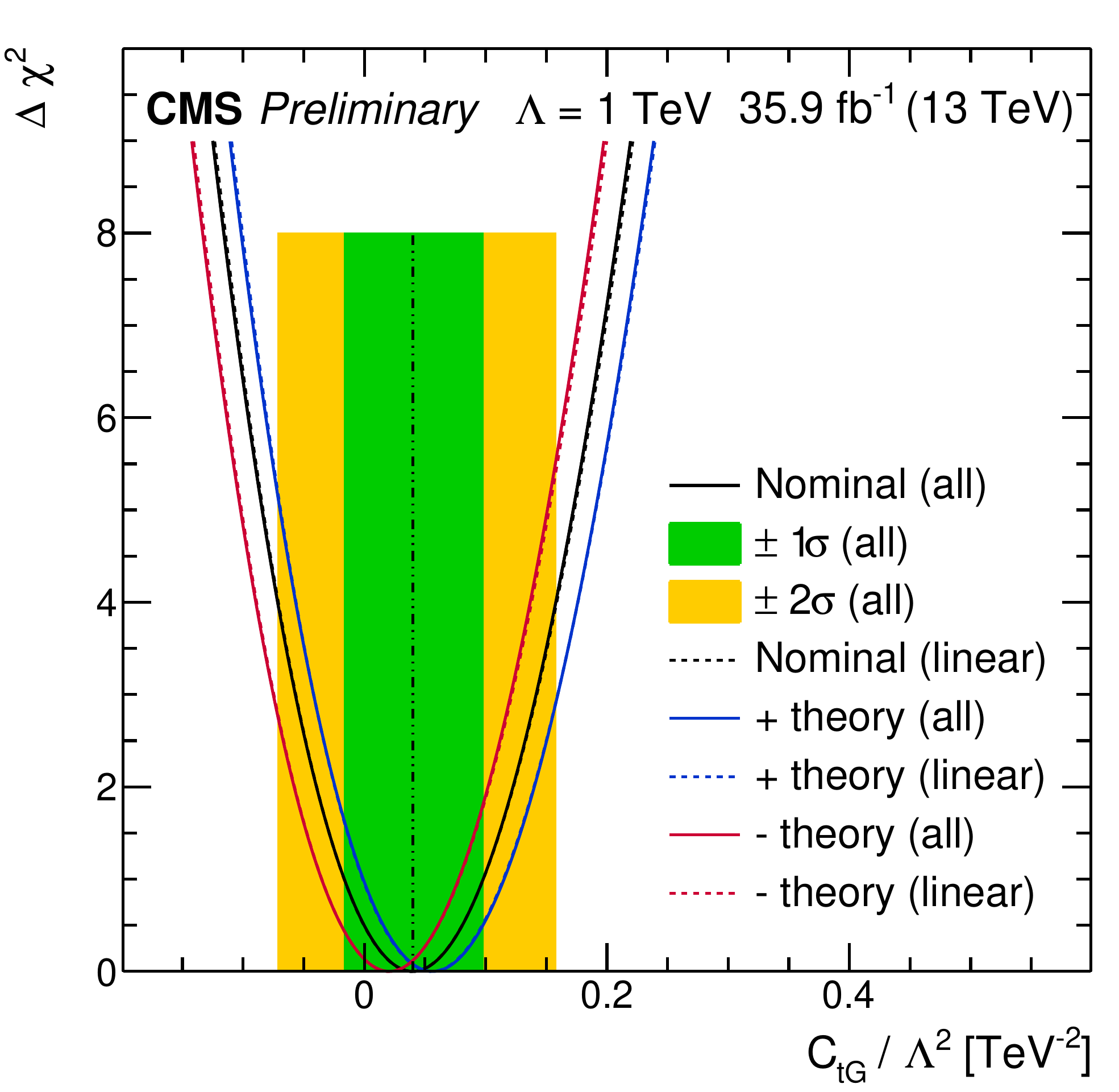}}
\end{minipage}
\caption[]{Left: systematic correlation matrix for all 132 bins measured in the CMS analysis~\cite{CMSresult}. Right: result~\cite{CMSresult} of the $\chi^{2}$ fit for \ctgl. 
The ``all'' result includes the full \ctgl-dependence of the \ttbar\ cross section, and the result when using
only the terms ``linear'' in \ctgl, which describe the interference with the SM, is almost the same.
}
\label{fig:interp}
\end{figure}

The measurement of the
\ttbar\ production spin density matrix is a powerful probe of the top
quark CMDM.
% and can be used to search for BSM phemomena. 
Using the measured distributions and covariance matrices~\cite{CMSresult}, 
and their predicted dependence~\cite{zhang} on the Wilson coefficient of the CMDM operator divided by the square of the new physics scale, \ctgl,
CMS performs a $\chi^{2}$ fit. The resulting constraints at 95\% CL, illustrated in \fig~\ref{fig:interp}, are $-0.07 < \ctgl\ < 0.16\:\mathrm{TeV}^{-2}$, the strongest constraints on \ctgl\ to date.

\subsection{``Top corridor'' SUSY \label{sec:susy} }

\begin{figure}
\begin{minipage}{0.575\linewidth}
\centerline{\includegraphics[width=1.0\linewidth]{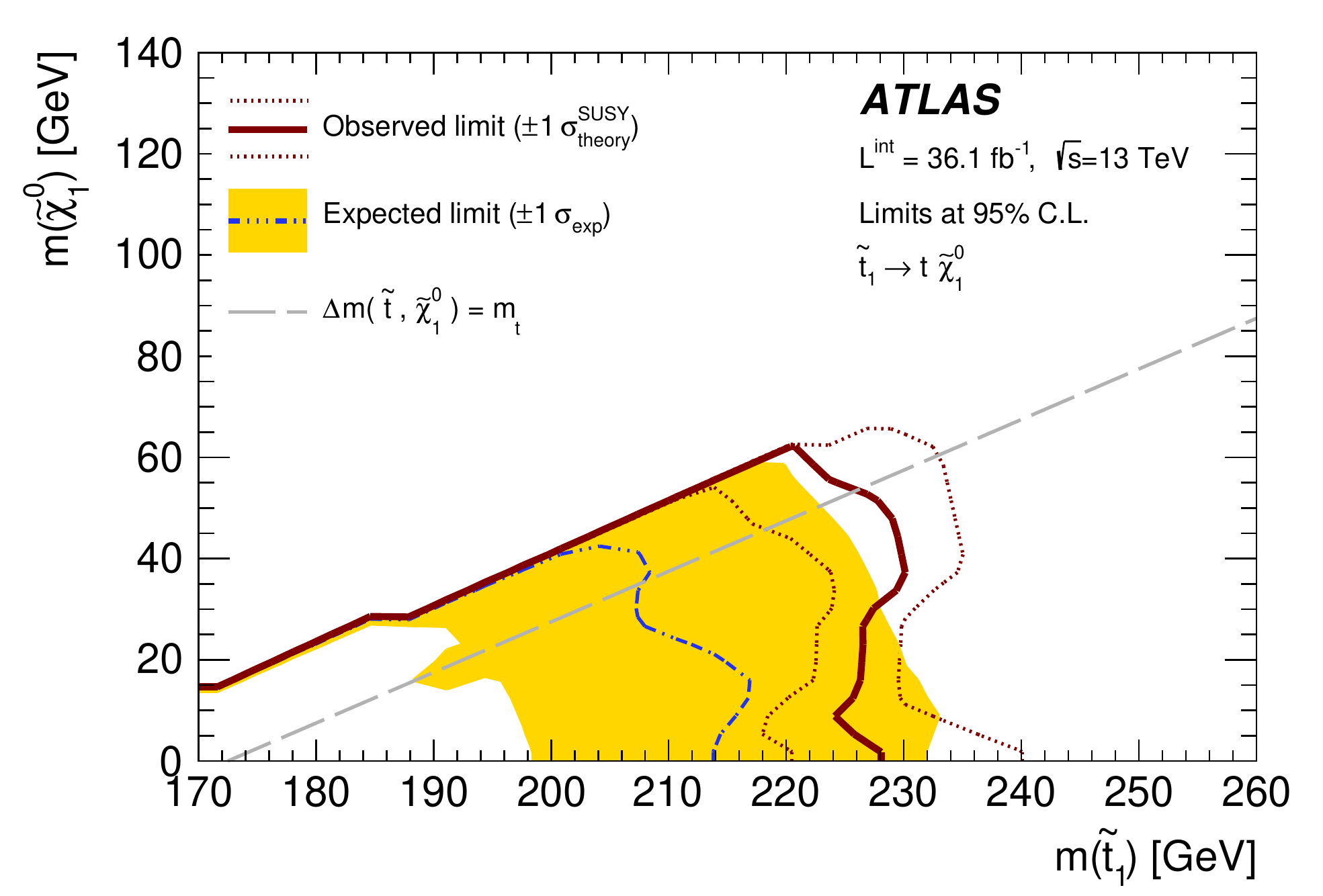}}
\end{minipage}
\hfill
\begin{minipage}{0.415\linewidth}
\centerline{\includegraphics[width=1.0\linewidth]{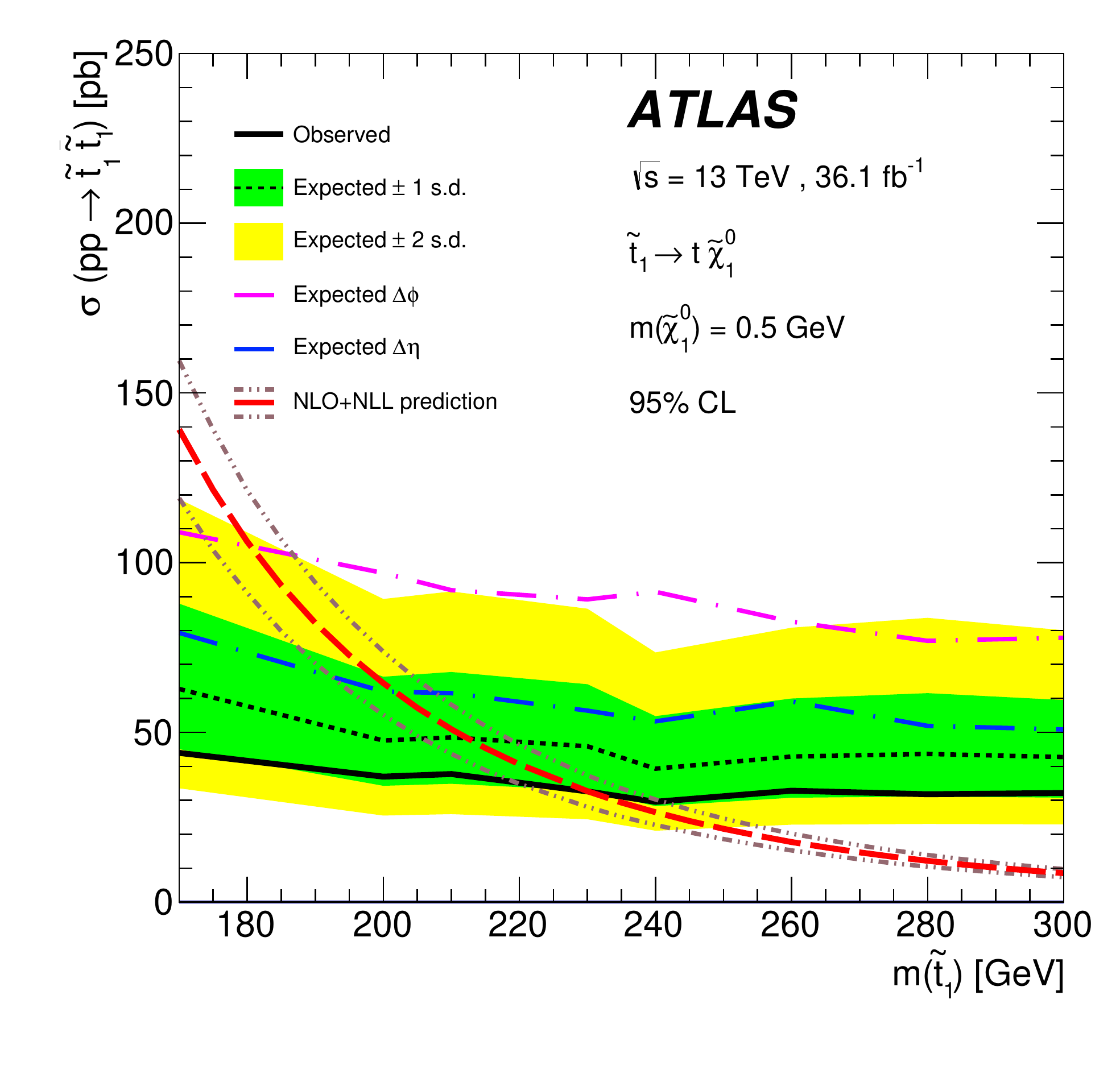}}
\end{minipage}
\caption[]{Left: observed and expected 95\% CL exclusion in the plane of $m_\mathrm{\tilde{t}}$ and $m_\mathrm{LSP}$~\cite{ATLASresult}. Right: limits on the $\tilde{t}\tilde{t^*}$ cross section at 95\% CL as a function of $m_\mathrm{\tilde{t}}$, assuming $m_\mathrm{LSP}=0.5\:\mathrm{GeV}$. The expected limits when using the \dphi\ and \deta\ distributions alone are shown by the magenta and blue dashed lines, respectively~\cite{ATLASresult}.}
\label{fig:interp2}
\end{figure}

Light top squarks ($\tilde{t}$) are favoured in natural supersymmetry (SUSY), and in many scenarios could decay to a top quark and the lightest SUSY particle (LSP), which would not interact with SM particles and would therefore escape the detector without producing a signal.
When the SUSY particle masses are such that the daughter top quark is produced almost at rest ($\Delta m = m_\mathrm{\tilde{t}} - m_\mathrm{LSP} = m_\mathrm{t}$), pair-produced $\tilde{t}\tilde{t^*}$ events can be difficult to distinguish from \ttbar\ events. Unlike in direct SUSY searches that rely on the presence of missing transverse momentum from the undetected LSP, in this scenario the main distinguishing characteristics are the spin correlations, which are absent in the scalar case, and the typically more central production kinematics of scalars, which translate to the separation in psuedorapidity of the leptons (\deta)~\cite{lightstop}. Using measured double-differential distributions in \dphi\ and \deta\ along with the total rate~\cite{ATLASresult}, ATLAS sets exclusion limits at 95\% CL (\fig~\ref{fig:interp2}, left). These results push the exclusion beyond those of existing direct searches in the region close to $\Delta m = m_\mathrm{t}$. The \deta\ information makes a greater contribution to the overall sensitivity than the \dphi\ information (see \fig~\ref{fig:interp2}, right).
The ATLAS analysis also unfolds the \deta\ distribution to the parton level (\fig~\ref{fig:indirect2}, right).
%(interestingly, this observable has minimal sensitivity to spin correlations)

\section{Summary}

Both ATLAS and CMS have presented measurements of \ttbar\ spin correlations using 
LHC $\mathrm{pp}$ collision data corresponding to an integrated luminosity of $36\:\mathrm{fb^{-1}}$ at $\sqrt{s}=13\:\mathrm{TeV}$.
The significant tension observed between the \dphi\ distributions measured by the ATLAS and CMS experiments and the SM predictions is likely explained by missing higher order corrections to the top quark kinematics, which become more important in the fiducial phase space accessible to the experiments.
The direct measurements of spin correlations are in good agreement with the SM predictions, and all spin-dependent coefficients of the \ttbar\ production density matrix have been probed for the first time at $\sqrt{s}=13\:\mathrm{TeV}$.
The spin correlation measurements are used to search for new physics in the form of a light top squark or an anomalous top quark
chromo-magnetic dipole moment, and stringent constraints are placed in both cases.
%the independent coefficients of the top quark spin-dependent parts of the \ttbar production density matrix

\section*{Acknowledgments}

This project has received funding from the European Union's Horizon 2020 research and innovation programme under the Marie Sk\l{}odowska-Curie grant agreement No. 752730.

%Copyright 2019 CERN for the benefit of the ATLAS and CMS Collaborations. CC-BY-4.0 license.

\footnotetext{Copyright 2019 CERN for the benefit of the ATLAS and CMS Collaborations. CC-BY-4.0 license.}

\end{document}